\documentclass[floatfix, preprint, showpacs, showkeys, preprintnumbers, nofootinbib, superscriptaddress]{revtex4-1}
\usepackage[utf8]{inputenc}
\usepackage[sort&compress]{natbib}
\usepackage{ulem}
\usepackage{bm}
\usepackage{times}
\usepackage{amssymb,amsbsy,amsmath,amsfonts}
\usepackage{graphicx}
\usepackage{float}
\usepackage{color}
\usepackage{morefloats}
\usepackage{rotating}
\usepackage{srcltx}
\usepackage{slashed}
\usepackage{subfigure}
\usepackage{multirow}
\usepackage{verbatim}
\usepackage{hyperref}
\usepackage{tabularx}

\newcommand{\RNum}[1]{\uppercase\expandafter{\romannumeral #1\relax}}

\begin{document}

\title{Novel Bayesian neural network based approach for nuclear charge radii}

\author{Xiao-Xu Dong}
\affiliation{School of Physics,  Beihang University, Beijing 102206, China}

\author{Rong An}
\affiliation{Key Laboratory of Beam Technology of Ministry of Education, Institute of Radiation Technology, Beijing Academy of Science and Technology, Beijing 100875, China}
\affiliation{Key Laboratory of Beam Technology of Ministry of Education, College of Nuclear Science and Technology, Beijing Normal University, Beijing 100875, China}

\author{Jun-Xu Lu}
\email[E-mail: ]{ljxwohool@buaa.edu.cn}
\affiliation{School of Space and Environment,  Beihang University, Beijing 102206, China}
\affiliation{School of Physics,  Beihang University, Beijing 102206, China}

\author{Li-Sheng Geng}
\email[E-mail: ]{lisheng.geng@buaa.edu.cn}
\affiliation{School of
Physics,  Beihang University, Beijing 102206, China}
\affiliation{Beijing Key Laboratory of Advanced Nuclear Materials and Physics, Beihang University, Beijing 102206, China }
\affiliation{School of Physics and Microelectronics, Zhengzhou University, Zhengzhou, Henan 450001, China }

\begin{abstract}

Charge radius is one of the most fundamental properties of a nucleus. However, a precise description of the evolution of charge radii along an isotopic chain is highly nontrivial, as reinforced by recent experimental measurements. In this paper, we propose a novel approach which combines a three-parameter formula and a Bayesian neural network. We find that the novel approach can describe the charge radii of all $A\ge40$ and $Z\ge20$ nuclei with a root-mean-square deviation about 0.015 fm. In particular, the charge radii of the calcium isotopic chain are reproduced very well, including the parabolic behavior and strong odd-even staggerings. We further test the approach for the potassium isotopes and show that it can describe well the experimental data  within uncertainties.

\end{abstract}


\maketitle

\section{Introduction}

Nuclear charge radius, as one of the most fundamental properties of a nucleus, plays a vital role in  our understanding of the complex dynamics of atomic nuclei and in showcasing various nuclear structure phenomena, such as neutron halo ~\cite{Nortershauser:2008vp}, shape coexistence and staggering~\cite{Yang:2016tyv,Marsh:2018wxs}, odd-even staggering~\cite{deGroote:2019yqm}, and nuclear magic numbers~\cite{Kreim:2013uqb,Gorges:2019wzy}. Lately, remarkable progress has been made in measuring the charge radii of those nuclei far from the $\beta$-stability line~\cite{Angeli:2013epw,Koszorus:2020mgn,Goodacre:2020sys,Li:2021fmk}. Although the global features of nuclear charge radii can be easily understood, e.g., $R\propto A^{1/3}$ or $Z^{1/3}$, with $Z$ and $A$ the proton  and mass numbers, there exist some fine structures that have eluded a complete understanding, e.g., the parabolic-like behavior and strong odd-even staggerings between $^{40}$Ca and $^{48}$Ca, and the abrupt increase from $^{48}$Ca to $^{52}$Ca, with the latter being a candidate of doubly magic nuclei~\cite{PhysRevC.31.2226,Wienholtz:2013nya,Rosenbusch:2015yma}. The charge radii of potassium isotopes, which have been recently measured~\cite{Koszorus:2020mgn,Kreim:2013uqb}, show   features below and above $N=28$ similar to those of calcium isotopes.

Many methods have been developed to predict nuclear charge radii, ranging from liquid drop models~\cite{Weizsacker:1935bkz,Brown:1984zz},  phenomenological parametrizations~\cite{Nerlo-Pomorska:1994dhg,Zhang:2001nt,Wang:2013zia,Sheng:2015poa}, sophisticated mean-field models~\cite{Geng:2005yu,Goriely:2016sdz,Pena-Arteaga:2016clz,Sarriguren:2019jfb}, to $ab$ $initio$ calculations with  chiral effective field theory interactions~\cite{Ekstrom:2015rta}. Most of these methods can describe the available data with a root-mean-square (RMS) deviation ranging from 0.07 to 0.02 fm. Nevertheless, none of them can provide satisfactory descriptions of the striking behavior of charge radii in calcium or potassium isotopes~\cite{GarciaRuiz:2016ohj,Koszorus:2020mgn}. Recently, by adding a semi-microscopic correction originating from the Cooper pair condensation, a modified relativistic mean field plus BCS (RMF(BCS)*) ansatz  has been proposed to describe the charge radii of the calcium~\cite{An:2020qgp} and potassium~\cite{An:2021rlw}
 isotopic chains.
 
In recent years, machine learning methods have found wide and successful applications in physics~\cite{Carleo:2019ptp,Bourilkov:2019yoi,Bedolla-Montiel:2020rio,Bedaque:2021bja}. In particular, Bayesian neural networks (BNNs), because of their ability to combine the strengths of artificial neural networks (ANNs) as ``universal approximators''~\cite{HORNIK1989359} and stochastic modeling, have been successfully applied to study various nuclear properties, such as masses~\cite{Utama:2015hva,Niu:2018csp}, incomplete fission yields~\cite{wang:2019pct}, charge yields of fission fragments~\cite{Qiao:2021mvr}, $\beta$-decay half-lives~\cite{Niu:2018trk}, nucleon axial form factor~\cite{Alvarez-Ruso:2018rdx}, proton radius~\cite{Graczyk:2014lba}, charge radii~\cite{Utama:2016tcl}, and nuclear liquid-gas phase transition \cite{Wang:2020tgb}.
In Ref.~\cite{Utama:2016tcl}, the proton number $Z$ and mass number $A$ of a nucleus are used as inputs to train a BNN, achieving a relatively good description of charge radii and reducing the RMS deviation by about 50\% in comparison with the underlying relativistic mean field (RMF) model. However, the BNN method fails to describe odd-even staggerings, in particular, those of calcium isotopes. Motivated by the success of Refs.~\cite{An:2020qgp,An:2021rlw},  we propose to improve the BNN method  using the so-called feature engineering technique to create two new input features, $\delta$ and $P$, from $Z$ and $A$. We will show that the refined BNN method can not only achieve a much improved description of experimental charge radii~\cite{Angeli:2013epw,Goodacre:2020sys,Li:2021fmk} but also can make reliable predictions for calcium and potassium isotopes with controlled uncertainties, which agree  well with the latest experimental data~\cite{Koszorus:2020mgn,Li:2021fmk}.

This article is organized as follows. In Sec. II, we construct the refined Bayesian neural network and explain how we categorize experimental charge radii for training and validation. Results and discussions are presented in Sec. III, followed by a short summary in Sec. IV.

\section{Theoretical Formalism}

Similarly to Ref.~\cite{Utama:2016tcl}, our purpose is to combine a theoretical model of charge radii and a Bayesian neural network to improve the description of nuclear charge radii. In our approach, the BNN is used to simulate the residuals between the theoretical predictions and the corresponding experimental data. In the following, we explain how to choose the theoretical model and how to construct the BNN.

As mentioned in the Introduction, a large number of microscopic and macroscopic models have been developed to describe atomic nuclei and most of them can provide reasonably good description of nuclear charge radii~\cite{Geng:2005yu,Utama:2016tcl,Li:2021fmk}. While the macroscopic models are much more convenient for practical applications, the microscopic models such as the RMF model~\cite{Geng:2005yu} are rather time consuming but contain more physics such as shell effects or pairing effects. However, in the present work, we prefer  to include these effects not in the underlying theoretical model but only via the inputs for the BNN to explore whether these effects can be learned by the BNN. Therefore,in this work, we choose the isospin-dependent NP formula\footnote{

We have checked that the hybrid approach consisting of the NP formula and the  BNN refinement can yield results similar to or even slightly better than the results one can achieve by replacing the NP formula with some microscopic models (for example,  the RMF model or the Weizs$\ddot{\mathrm{a}}$cker Skyrme~(WS$^{*}$) model~\cite{Li:2021fmk}).

} developed  by  Nerlo-Pomorska and Pomorski~\cite{Nerlo-Pomorska:1994dhg}:
\begin{align}
R_{\mathrm{NP}}(Z,A)=r_A A^{\frac{1}{3}}\left[ 1-b(\frac{N-Z}{A})+\frac{c}{A}\right],
\end{align}
where $r_A$ = 0.966 fm, b = 0.182, and c = 1.652~\cite{Bayram:2013jua}.
This formula can describe nuclear charge radii at a level similar to the more sophisticated relativistic mean field model~\cite{Utama:2016tcl}.

\begin{figure}
    \centering
    \includegraphics[angle=270,width=0.8\textwidth]{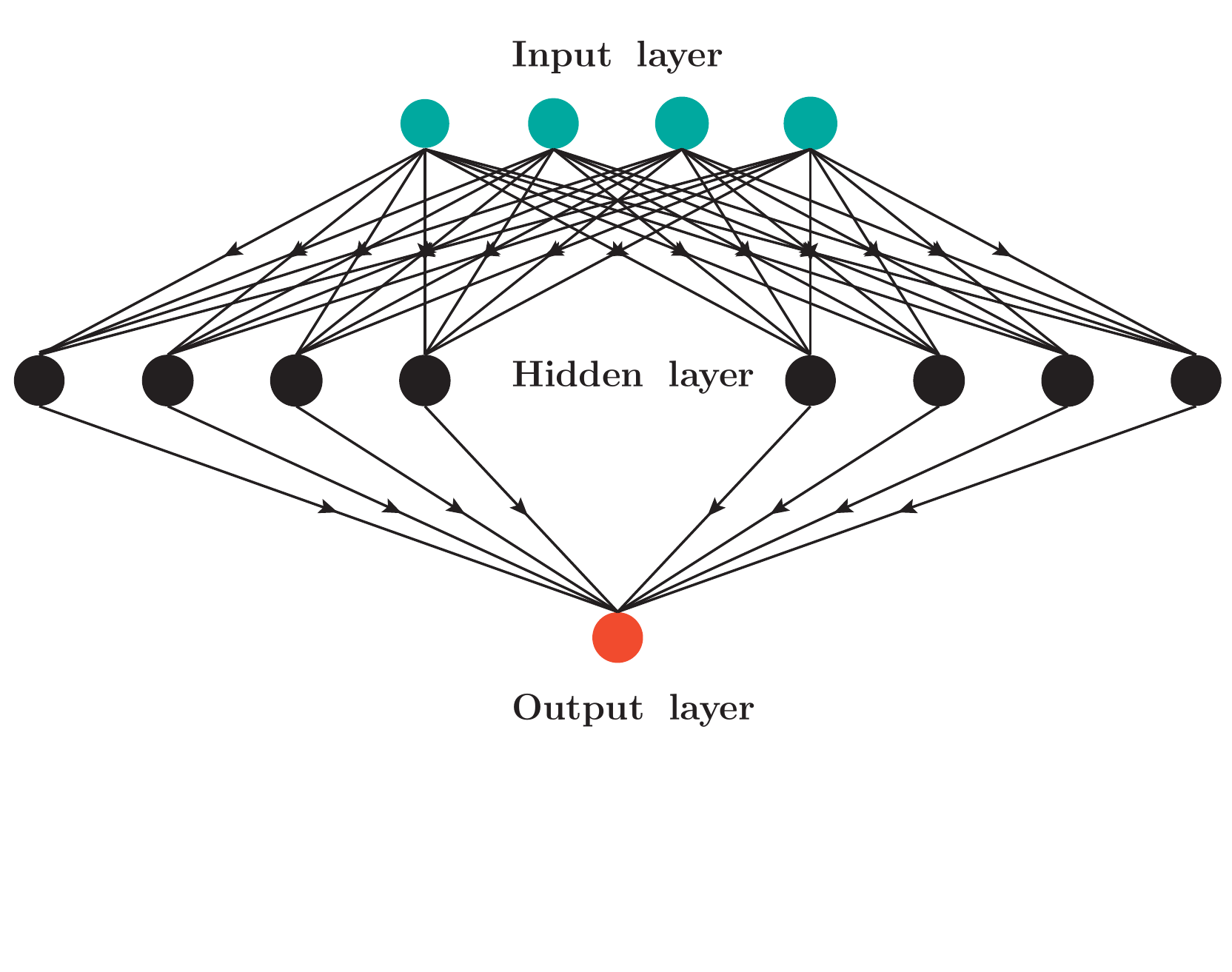}
    \caption{ 
    Structure of the artificial neural network used in this work. The number of neurons in the input layer is either 2 or 4. The number of hidden layers is 1 and the number of neurons in the hidden layer is $40$. The number of neurons in the output layer is $1$.
    } 
    \label{Structure}
\end{figure}

There are two main components in the BNN~\cite{Neal}: one is the artificial neural network and the other is the Bayesian inference system. As shown in Fig.~\ref{Structure}, the artificial neural network we use is a fully connected feed-forward artificial neural network with one hidden layer. Mathematically, it has the following form:
\begin{align}
f(x,\omega)=a+\sum\limits_{j=1}\limits^H b_j \tanh(c_j+\sum\limits_{i=1}\limits^I d_{ji}x_i),\label{f}
\end{align}
where the parameters of the neural network are $\omega=({a, b_j, c_j, d_{ji}}$), $I$ is the number of input layer neurons, $H$ is the number of hidden layer neurons, and $x$ is the set of inputs $x_i$. The function in Eq.~\eqref{f} contains $1+H(2+I)$ parameters.

The Bayesian inference is based on Bayes's theorem, which reads
\begin{align}
p(\omega|x,t)=\frac{p(\omega)p(x,t|\omega)}{p(x,t)},
\end{align}
where $p(\omega)$ is the prior probability of the parameters of the neural network, $p(x,t|\omega)$ is the likelihood based on the actual data,  $p(\omega|x,t)$ is the posterior probability calculated from the prior probability and likelihood, which is used to predict the unknown data, $p(x,t)$ is the marginal likelihood, and $t$ is the set of target data $t_i$. Generally speaking, the prior probability encodes  our knowledge on the subject under study. In the present case, we assume that the four sets of parameters $\omega=({a, b, c, d})$ are independent of each other and each of them obeys a Gaussian distribution centered around 0 with a width controlled by a hyperparameter. As shown in Ref.~\cite{Neal}, the ``gamma" probability  distribution is used for the hyperparameter. Similarly, a Gaussian distribution is used for the likelihood in terms of an objective  function obtained from a least-squares fit to the training data:
\begin{align}
p(x,t|\omega)=\exp(-\frac{\chi^2(\omega)}{2}),
\end{align}
with
\begin{align}
\chi^2(\omega)=\sum\limits_{i=1}\limits^N\left(\frac{t_i-f(x_i,\omega)}{\Delta t_i}\right)^2,
\end{align}
where $N$ is the number of training data. The noise error $\Delta t_i$ is an important quantity in the Bayesian method. However, it was usually simplified by taking a fixed value or sampling from a prior distribution~\cite{Niu:2018csp}, both of which do not contain much physics. In this work, we use the experimental uncertainties of charge radii as noise errors.

Differently from Ref.~\cite{Utama:2016tcl}, where only the proton number and mass number of a nucleus are used as inputs to predict its charge radius, we propose to enlarge the number of inputs using feature engineering. Such a method could be referred to as a physically motivated BNN method. It is well known that for many isotopes, such as the calcium and potassium isotopes, charge radii exhibit strong odd-even staggerings. In Refs.~\cite{An:2020qgp,An:2021rlw}, such staggerings are related to the so-called Cooper pair condensation or pairing interaction. Inspired by the successful description of calcium~\cite{An:2020qgp} and potassium~\cite{An:2021rlw} charge radii, we construct from $Z$ and $N$ two more inputs, i.e., $\delta$ and $P$, which are defined as
\begin{align}
\delta&=\frac{(-1)^Z+(-1)^N}{2},\\
P&=\frac{\nu_p\nu_n}{\nu_p+\nu_n}.
\end{align}
 The pairing term $\delta$ is related to  nuclear pairing effects and the promiscuity factor $P$~\cite{Casten:1987zz,Casten_1996} is related to shell closure effects. In the definition of $P$,  $\nu_{p(n)}$ is the difference between the proton (neutron) number of a particular nucleus and the nearest magic number. In this work, the neutron and proton magic numbers are taken as $Z=8, 20, 28, 50, 82, 126$ and $N=8, 20, 28, 50, 82, 126, 184$~\cite{KIRSON200829}.

With these two more inputs, the input data for the refined BNN model are $x\equiv (Z,A,\delta,P)$.
The target data set $t$ values are $\delta R_{\mathrm{ch}}=R_{\mathrm{exp.}}-R_{\mathrm{NP}}$, i.e., the residuals between experimental data and theoretical predictions of charge radii given by the NP formula.

Unlike other artificial neural networks, the parameters of BNN after training obey a posterior probability. Therefore, the Bayesian predictions for the target data are:
\begin{align}
\langle f_n\rangle=\int f(x_n,\omega)p(\omega|x,t)d\omega=\frac{1}{K}\sum\limits_{k=1}\limits^K f(x_n,\omega_k),\label{fn}
\end{align}
where $\langle f_n \rangle$ are the Bayesian predictions, $x_n=(Z_n,A_n,\delta_n,P_n)$ are the input data, $f(x_n,\omega_k)$ are the neural network predictions for $\delta R_{\mathrm{ch}}(Z_n,A_n)$ for a given set of parameters $\omega_k$, and $K$  is the total number of effective samples. In this work, we use the Markov chain Monte Carlo (MCMC)  method~\cite{Neal}  to obtain the Bayesian predictions.  As long as the MCMC method is used, the marginal likelihood $p(x,t)$ can be ignored. A distinct advantage of BNNs is that they can provide a proper estimate  of confidence intervals (CIs) :
\begin{align}
\Delta=\sqrt{\langle f_n^2 \rangle-\langle f_n \rangle^2},
\end{align}
where  $\langle f_n^2 \rangle $ is obtained following the same procedure as in obtaining $\langle f_n\rangle $.

\section{Results and discussions}
In this work, inspired by Refs.~\cite{An:2020qgp,An:2021rlw}, we propose a physically motivated BNN model to study nuclear charge radii and  to check whether one can reproduce some of the known fine structure and make reliable predictions.

For very light nuclei, because of their small masses and large fluctuations in charge distributions, it is often argued  that regarding  charge  radii as  bulk properties is of little meaning~\cite{Zhang:2001nt,Sheng:2015poa}. Therefore, we only study those nuclei with $A\ge40$ and $Z\ge20$. For the training set, we include the experimental data given in Ref.~\cite{Angeli:2013epw}, which consist of in total 820 data. The more recent experimental data ~\cite{Goodacre:2020sys,Li:2021fmk}, containing 113 data for nuclei with $A\ge40$ and $Z\ge20$, are used as the validation set to test the  predictive power  of our BNN model. The entire set combines the training and validation sets and contains 933 data of nuclear charge radii.

For the sake of easy reference, we use  ``D4'' to denote  the model combining the NP formula and the four-input neurons BNN, and ``D2''  the model combining the NP formula and the two-input neurons BNN. It should be noted that the D2 model is similar to one version of the BNN models of Ref.~\cite{Utama:2016tcl}.
Since the input data and target data have been determined, we can use them to train the BNN. To quantify the extent of the BNN refinement of the NP formula, we compute the RMS deviation between the D4(D2) outputs and experimental data:
\begin{align}
\sigma_{v}=\sqrt{\frac{1}{N_v}\sum\limits_{i=1}\limits^{N_v}\left(R_{i}^{(\mathrm{exp.})}-R_{i}^{(\mathrm{theo.})}\right)^2}
\end{align}
where $N_v$ is the total number of charge radii  used in the validation set. The RMS deviations of the training set and the entire set can be calculated in the same way. The corresponding results for the NP formula, D2, and D4 models are displayed in Table~\ref{rms} for the three data sets, i.e., the training set, the validation set, and the entire set.

\begin{table}[htpb]
\caption{RMS deviations of charge radii predicted by the NP formula, D2 and D4 models.}
\begin{center}
\begin{tabular}{c c c c }
  \hline\hline
  & Training set &Validation set& Entire set\\
  \hline
  $\sigma^{(\mathrm{D2})}~(\mathrm{fm})$ & 0.0161 & 0.0250  &0.0174\\
  $\sigma^{(\mathrm{D4})}~(\mathrm{fm})$ &0.0143  & 0.0187  &0.0149\\
  $\sigma^{(\mathrm{NP})}~(\mathrm{fm})$&0.0394 &0.0300   &0.0384\\
  \hline
  $\frac{\sigma^{(\mathrm{NP})}-\sigma^{(\mathrm{D2})}}{\sigma^{(\mathrm{NP})}}$& 0.59  &0.16 & 0.55\\
  $\frac{\sigma^{(\mathrm{NP})}-\sigma^{(\mathrm{D4})}}{\sigma^{(\mathrm{NP})}}$& 0.64  & 0.38&0.61\\

  \hline\hline

\end{tabular}
\label{rms}
\end{center}
\end{table}

 As expected, the D4 model achieves the least RMS deviation for the training set, which is only about 36\% of that  of the NP formula. Similar improvement is also found for the D2 model.  For the validation set, the RMS deviation of the D4 model is only larger by about 31\%, while that of the D2 model increases by 55\%. Interestingly and somehow unexpectedly, the NP formula describes the validation set better than the training set.  We note in passing that, in Ref.~\cite{Sheng:2015poa}, the modified NP formula which uses $\delta$ and $P$ as two more degrees of freedom achieves an RMS deviation  of about 0.0223 fm, which should be compared with 0.0143 fm achieved by our D4 model.

 \begin{figure}[ht]
 \centering
 \includegraphics[width=1\textwidth]{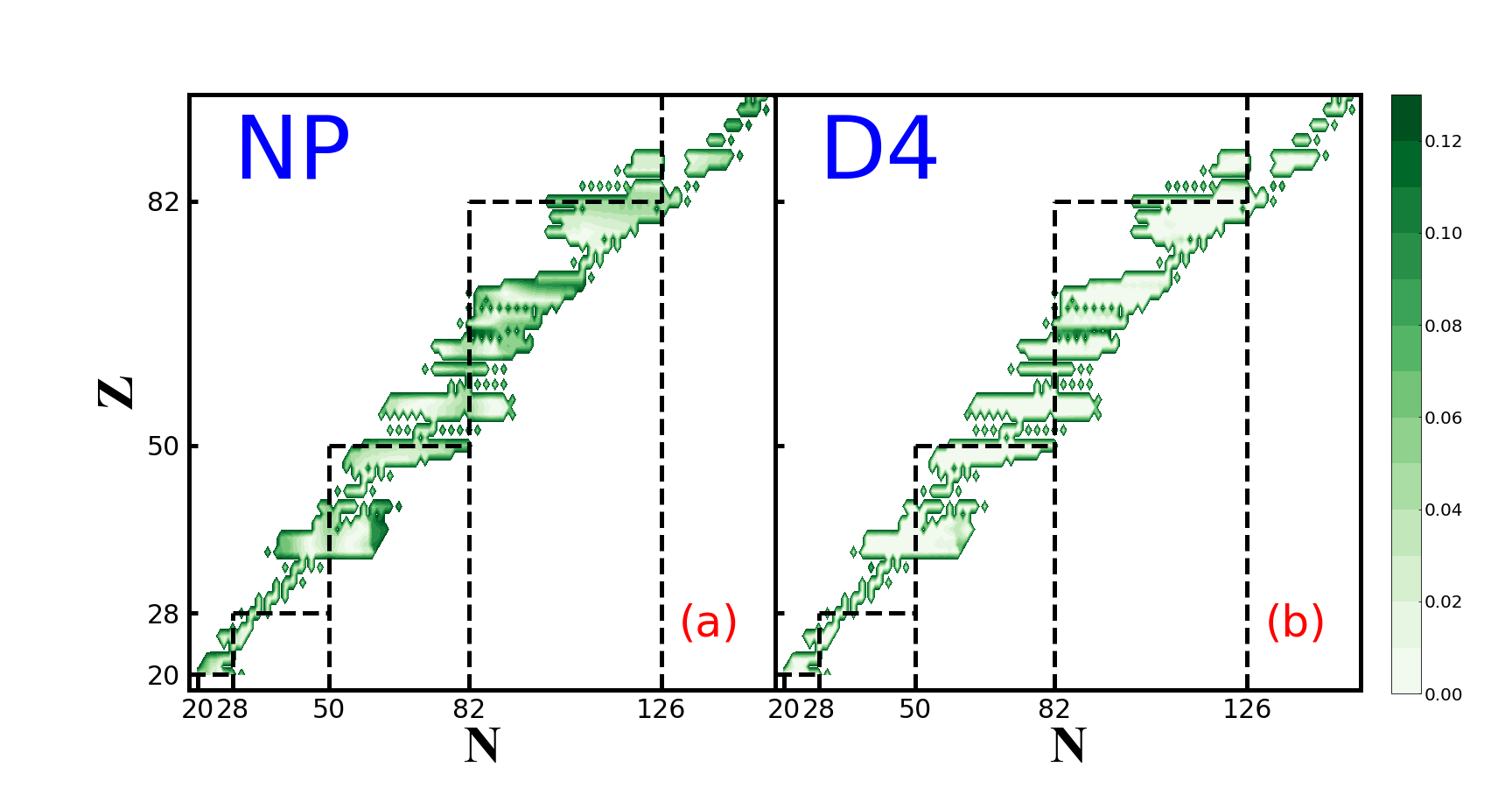}
 \centering
 \caption{ Predictions for the residuals (in units of fm) between the experimental charge radii and  the NP formula~\cite{Nerlo-Pomorska:1994dhg,Bayram:2013jua} (a) and the D4 model (b) predictions.}\label{sum}
 \end{figure}

For the charge radii of all the  nuclei in the entire set, the differences between the theoretical results of the NP formula (the D4 model) and the experimental data  are illustrated in Fig.~\ref{sum}. It can be clearly seen that in comparison with the NP formula, the D4 model improves greatly the description of nuclear charge radii.

\begin{figure}[htbp]
\centering
\subfigure{
\begin{minipage}[t]{0.49\linewidth}
\centering
\includegraphics[width=1.0\textwidth]{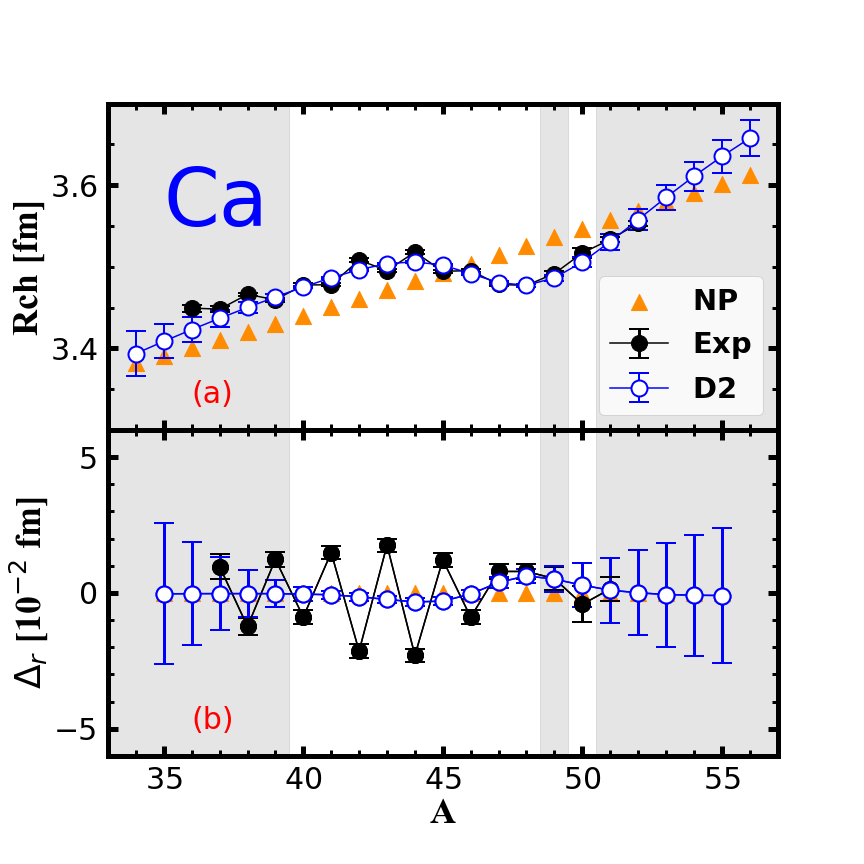}
\end{minipage}%
}%
\subfigure{
\begin{minipage}[t]{0.49\linewidth}
\centering
\includegraphics[width=1.0\textwidth]{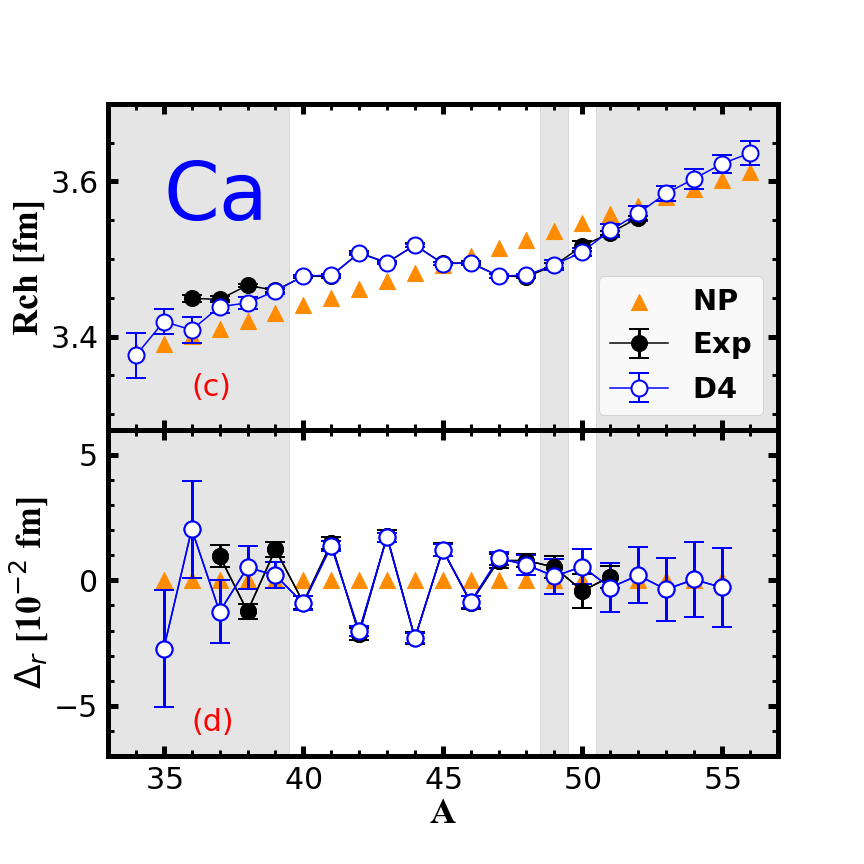}
\end{minipage}%
}%
\\

\centering
\caption{Charge radii (a, c) and $\Delta_r$ (b, d) of calcium isotopes predicted by the  NP formula~\cite{Nerlo-Pomorska:1994dhg,Bayram:2013jua}, D2 and D4 models, in comparison with the experimental data~\cite{Angeli:2013epw,Li:2021fmk}. The data in the grey area are predictions, i.e., they are not contained in the training set. }\label{calcium}
\end{figure}

Although the BNN models with 4 and 2 input neurons seem to describe the training set at a similar level, because of the extra information, ``features'' ($\delta$ and $P$), contained in the D4 model, it is expected to better describe  some fine structure of charge radii, e.g., the odd-even staggerings of charge radii of calcium isotopes. This is indeed the case, as shown in  Fig.~\ref{calcium}. There are several features about the charge radii of calcium isotopes which pose great challenges to current theoretical models, as summarized in Ref.~\cite{An:2020qgp}. First, the charge radii of $^{40}$Ca and $^{48}$Ca  are quite close to each other. Second, the odd-even staggerings are very strong in the $A<48$ region and an unexpected large increase is observed in  the $A>48$ region~\cite{Angeli:2013epw,GarciaRuiz:2016ohj,Miller2019}.

As can be seen from Fig.~\ref{calcium}, the dependence of the charge radii on the mass number predicted by the NP formula is almost linear, which fails to give a  satisfactory description  of the experimental data. With  two-input neurons  $(Z,A)$, the D2 model describes much better the experimental data but still fails to reproduce the large odd-even staggering, as found by Utama et al.~\cite{Utama:2016tcl}. With four-input neurons ($Z,A,\delta,P)$, the D4 model describes the data much better, in particular, the odd-even effects of  the training set ($40\le A\le 48$). Both the D2 and D4 models predict a large increase of  the charge radii in the ~$A\ge49$ region. The predicted charge radii for $^{49,51,52}$Ca are in good agreement with the experimental data, which demonstrates the predictive power of artificial neural networks.  Nevertheless, we note that the predicted odd-even staggerings by the D4 model for the nuclei with $36\le A \le39$  are a bit stronger than the data. 

In the lower panels of Fig.~\ref{calcium}, we compare
the odd-even staggerings $\Delta_r$ defined by
\begin{align}
    \Delta_r(N,Z)&=\frac{1}{2}\left[ R(N-1,Z)-2R(N,Z)+R(N+1,Z)\right],
\end{align}
where $R(N,Z)$ is the RMS charge radius for a nucleus with neutron number $N$ and proton number $Z$. It is clear that neither the NP results nor the D2 results show any odd-even staggerings while the D4 results are in perfect agreement with data. The predictions shown in the grey area, however, suffer from relatively large uncertainties.


\begin{figure}[htbp]
\centering
\subfigure{
\begin{minipage}[t]{0.49\linewidth}
\centering
\includegraphics[width=1.0\textwidth]{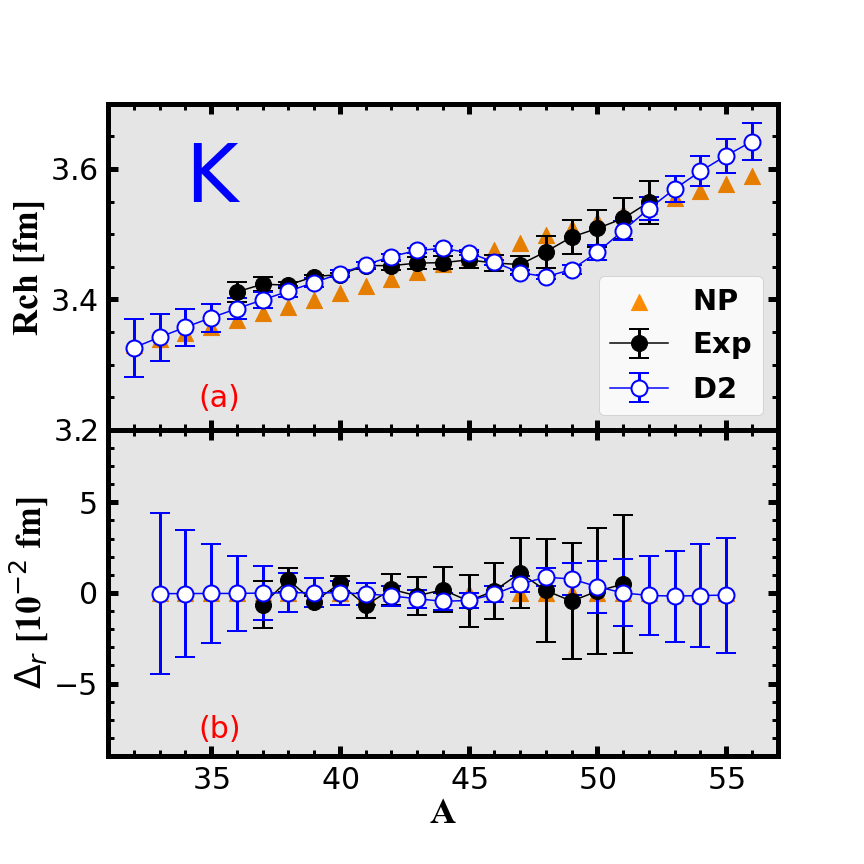}
\end{minipage}%
}%
\subfigure{
\begin{minipage}[t]{0.49\linewidth}
\centering
\includegraphics[width=1.0\textwidth]{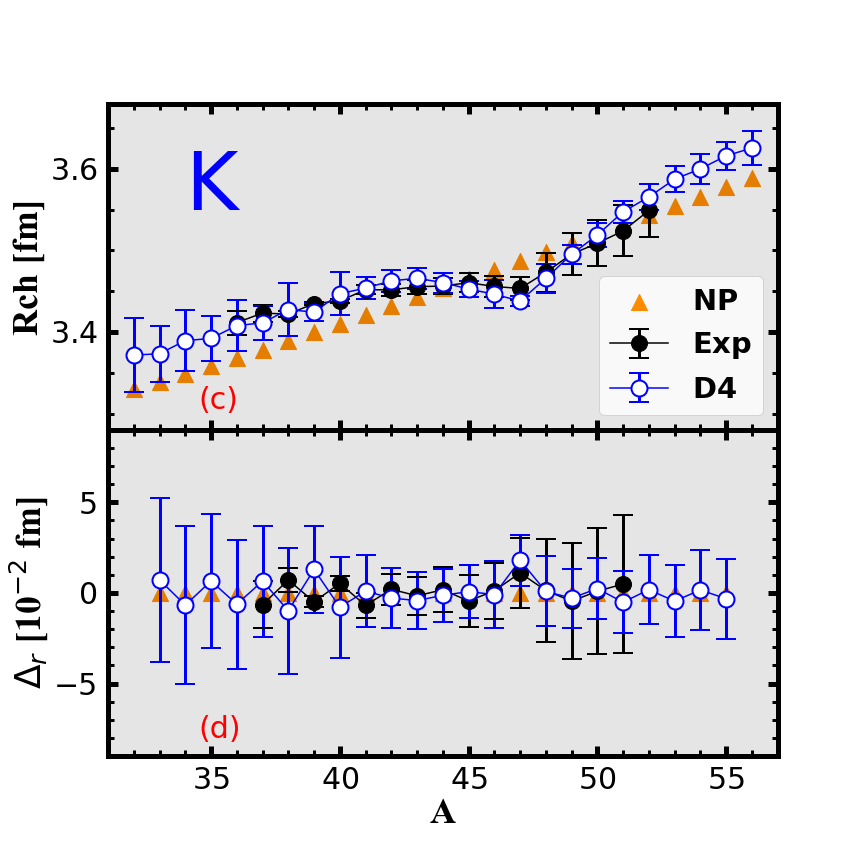}
\end{minipage}%
}%
\centering
\caption{Charge radii (a, c) and $\Delta_r$ (b, d) of the potassium isotopes predicted by the  NP formula~\cite{Nerlo-Pomorska:1994dhg,Bayram:2013jua}, D2 and D4 models, in comparison with the experimental 
data~\cite{Angeli:2013epw,Koszorus:2020mgn,Li:2021fmk}.}\label{potassium}
\end{figure}

It should be noted that the results for calcium isotopes are not completely predictions because the charge radii of some isotopes are used already for training the BNN model. In order to further verify the predictive power of the D2 and D4 models,  we calculate the charge radii of potassium isotopes, none of which are used in the training process of our BNN models. The results are shown in Fig.~\ref{potassium}. Somehow, surprisingly, the predictions of the D4 model are in very good agreement with the experimental data. It can be seen  from Fig.~\ref{potassium} that most experimental data are within the uncertainties of the D4 model, which implies that the charge radii predictions and corresponding uncertainties given by the D4 model are quite reasonable.

Similar to the calcium isotopic chain, neither the  NP predictions nor the D2 predictions show any odd-even staggerings while the corresponding results of the D4 model show strong odd-even staggerings, which agree with the experimental data.


Recently, a naive Bayesian probability (NBP) classifier was employed to study nuclear charge radii by fitting to residuals between experimental data and theoretical predictions  by the Skyrme-Hartree-Fock-Bogoliubov model and Sheng's formula~\cite{Ma:2020rdk}. The accuracy of the  theoretical predictions for charge radii is improved by about 20\% for the validation set after the NBP refinement. However, it cannot reproduce  the odd-even staggerings of charge radii of the calcium isotopes.

In Ref.~\cite{Wu:2020bao} a feed-forward artificial neural network (ANN) was adopted to directly fit to the experimental data of charge radii. The trained ANN approach reproduces well not only the kinks of charge radii at $N=82$ and $126$ but also the charge radii of calcium isotopes with even neutron number. However, the entire set only contains  347  data, far from enough to make reliable extrapolations for nuclear charge radii throughout the nuclear chart.

\section{Summary and outlook}

We built a hybrid model which combines a three-parameter parametrization and the flexibility of a Bayesian neural network (BNN) to study nuclear charge radii. We show that with physically motivated features, i.e., pairing and shell effects, one can achieve an unprecedented description of nuclear charge radii. Compared to the three-parameter parametrization, the RMS deviation achieved by the hybrid model is lower by nearly 40\%. In particular, the strong odd-even staggerings of calcium isotopes are described very well. In addition, the predictions of the hybrid approach are shown to be in very good agreement with data for potassium isotopes. Another advantage of the hybrid approach is that it can give an estimate of theoretical uncertainties.

The comparison with two- and four-input neurons demonstrated that providing more physical information to the BNN is crucial for the success of the hybrid approach. This should be explored for scenarios where data are limited, which is often the case in nuclear physics.

\section{Acknowledgments}
This work was partly supported by the National Natural Science Foundation of China (NSFC) under Grants No. 11975041, No. 11735003, No. 12105006, and No. 11961141004. R.A. is supported in part by the Reform and Development Project of the Beijing Academy of Science and Technology under Grant No. 13001-2110.

\bibliography{BNN.bib}

\end{document}